\documentclass[12pt,preprint]{aastex}
\shorttitle{A Compact ESE Toward AO~0235+164}
\shortauthors{C. E. Senkbeil et al.}
\bibpunct{(}{)}{;}{a}{,}{,}
\begin{document}
\bibliographystyle{plainnat}


\title{A Compact Extreme Scattering Event Cloud Towards AO~0235+164}
\author{C. E. Senkbeil, S. P. Ellingsen, and  J. E. J. Lovell}\affil{School of Mathematics and Physics, Private Bag 37, University of Tasmania, Hobart, TAS 7001 , Australia}

\and
\author{J.-P. Macquart$^\dagger$}
\affil{NRAO, P.O. Box 0, Socorro, NM 87801, U.S.A. and Dept. of Astronomy, California Institute of Technology, Pasadena, CA 91125, U.S.A.}\altaffiltext{$\dagger$}{NRAO Jansky Fellow}
\and
\author{G. Cim\`o}
\affil{Joint Institute for VLBI in Europe, Postbus 2, 7990 AA Dwingeloo, Netherlands}
\and
\author{D. L. Jauncey}
\affil{CSIRO Australia Telescope National Facility, Epping New South Wales, Australia}

\begin{abstract}
We present observations of a rare, rapid, high amplitude Extreme
Scattering Event toward the compact BL-Lac AO~0235+164 at 6.65\,GHz.  The ESE cloud is compact; we estimate its diameter between 0.09 and 0.9\,AU, and is at a distance of less than $3.6$\,kpc.  Limits on the
angular extent of the ESE cloud imply a minimum cloud electron density of
$\sim 4 \times 10^3$\,cm$^{-3}$. Based on the amplitude and timescale of the ESE observed here, we suggest that at least one of the  transients reported by \citet{bower2007} may be attributed to ESEs.
\end{abstract}


\section{Introduction}

The $z=0.940$ \citep{cohen1987} flat spectrum BL Lac AO~0235+164 has
been known to exhibit variability since its discovery \citep{macleod1976}
 over a broad range of wavelengths and timescales (\cite{raiteri2006} and references therein). Two absorbing galaxy complexes exist along the line of sight to this source at
$z=0.524$ and $z=0.851$ \citep{burbidge1976,rieke1976}, hindering interpretation of
the spectrum and possibly contributing to the observed variability
(e.g. by microlensing by stars in the foreground system
\citep{ostriker1985}).  The source is amongst the most compact radio
AGN: it exhibits long-term variability down to meter-wavelengths and
is slightly resolved by high angular resolution VLBI at 43\,GHz
\citep{frey2000,piner2006}.

The extreme compactness of this source renders the interpretation of
its variability difficult.  For instance, there is debate whether the
origin of the centimeter-wavelength intra-day variability in observed
AO~0235$+$164 is primarily intrinsic or due to interstellar
scintillation (ISS) \citep{kraus1999,lovell2003}, and on what
timescales each contributes.  In addition, the source compactness
renders it highly susceptible to flux density deviations by the compact
intervening refracting structures in the local ISM that give rise to
Extreme Scattering Events (ESEs) \citep{fiedleretal87} because it is
easy for even a small cloud to subtend the angular extent of such a
small source.

Of the three main causes of cm-wavelength variability, ESEs are the
most rare and certainly the least understood.  Almost all
flat-spectrum AGN show variability \citep{altschuler1977}. The
MASIV survey found that $\sim 20$\% of this population exhibits
scintillation at any one time with over 56\% exhibiting ISS at any
stage over the course of a year \citep{lov2007,jauncey2007}.  However, the estimated rate of ESE
events in compact AGN is only 0.013\,source$^{-1}$yr$^{-1}$ \citep{fiedleretal87}

There are various plausible interpretations as to the origins of the
structures that give rise to ESEs.  These include intrinsically
turbulent ionized clouds, purely refractive (gaussian) lenses, and
primarily neutral clouds enveloped by a thin ionized sheath that
responsible for the cloud's refractive properties
\citep{fiedleretal94,romani87,clegg98,ww98}.  These models pose a number
of challenges; the electron densities implied by the plasma lens
models requires the clouds to be $\sim 10^3$ times overpressured with
respect to the ambient ISM, while the neutral-cloud interpretation
implies that such structures would contain a large fraction of the
baryonic dark matter content of the Milky Way.

In this paper we report an unusually short-timescale ESE in
AO~0235$+$164.  Our observations are reported in section 2, while in
section 3 we show that both intrinsic variability and interstellar
scintillation are incapable of explaining this event, and discuss the
physical properties of the ESE necessary to give rise to the observed
lightcurve.  Our conclusions are presented in section 4.

\section{Observations}\label{sec:obs}
AO~0235+164 has been monitored quasi-continuously since 2003 as part
of the Continuous Single-dish Monitoring of Intraday variability at
Ceduna (COSMIC) program \citep{mcculloch2005}.  The
observations were conducted with the University of Tasmania Ceduna 30-m antenna at a center
frequency of 6.65~GHz with a 300\,MHz bandwidth. The flux density was
sampled by scanning across the source in forward and reverse directions in both Right Ascension and Declination.  The flux density measurements are obtained from
the height of the gaussian profile of the source signal above the
system temperature baseline. All the scans are scaled to a noise diode, which in turn is calibrated against 3C~227, which at
this frequency has a flux density of 1.99\,Jy  \citep{baars1977}. Measured pointing offsets are used to correct the amplitude in orthogonal scans to
minimise the impact of inaccurate pointing on the measured
amplitude. A correction is made to account for gain dependence of the antenna due to
the distortion with respect to elevation. The four
fitted amplitudes are averaged together to constitute a single calibrated sample.

\begin{figure}
\epsscale{1.3}
\centerline{\plotone{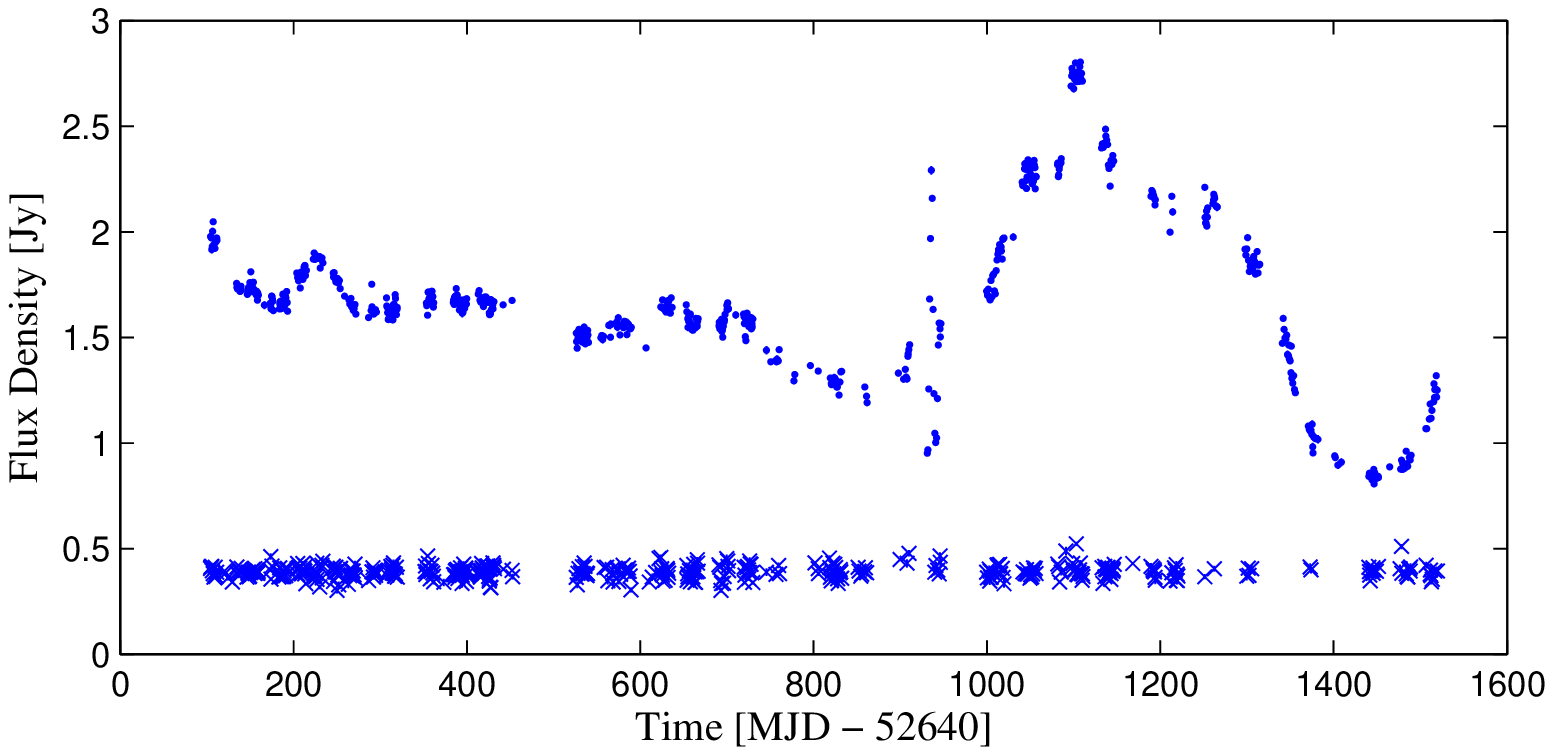}}
\caption{The 6.65\,GHz lightcurve of AO~0235$+$164 (points) since 2003 Jan 1 (MJD 52640). The corresponding lightcurve of the calibrator, 3C~277, is shown (crosses), with 1.6\,Jy subtracted from its flux densities for display purposes. The averaging time for each flux density sample is 12 hours.\label{fig:entirelc}}
\end{figure}

The flux density monitoring data from Ceduna are known to suffer variations that are systematic in nature  and predominantly diurnal in timescale, being related to changes in air temperature at the observatory. These errors scale with source flux density and affect source and calibrator equally (as they are approximately the same brightness). The amplitude of this systematic effect for the present case is 3\%, a factor of $\sim 40$ times smaller than the variability in AO~0235+164 discussed here. The calibrator data shown in Figure~\ref{fig:entirelc} demonstrates that this effect is negligible. Moreover, the variations occur on a much longer timescale than the diurnal systematic errors.

\begin{figure*}
\epsscale{1.4}
\centerline{\plotone{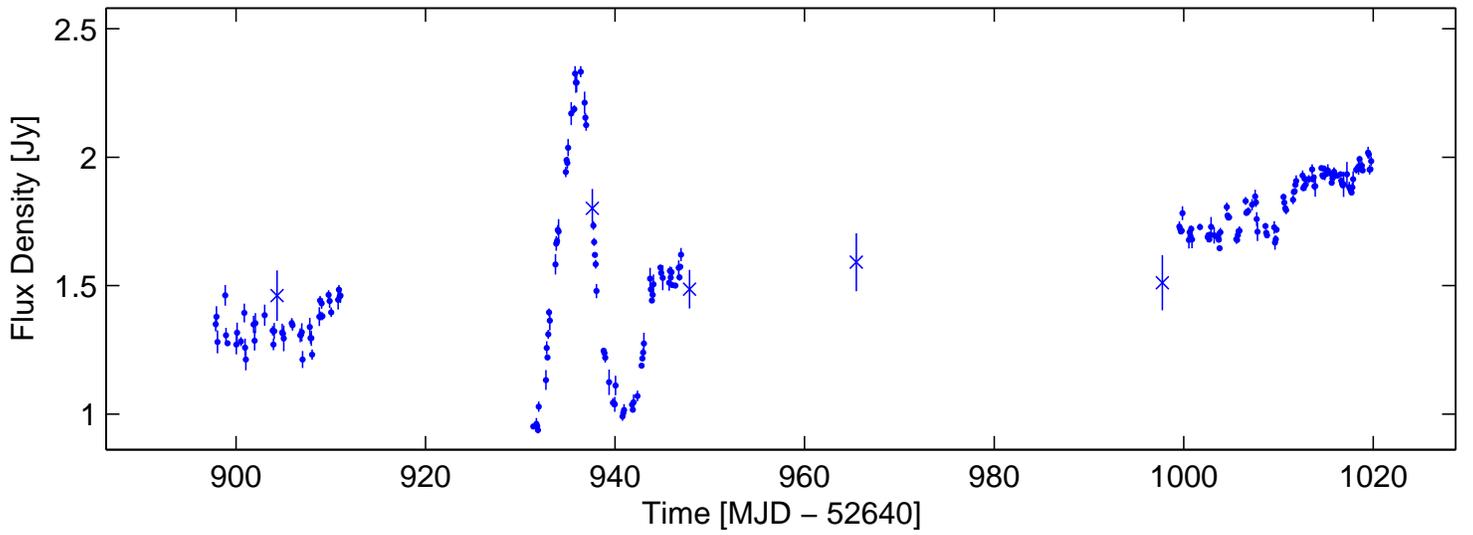}}
\caption{The 6.65\,GHz lightcurve of AO~0235+164 during the rapid flux density variation in 2005, The averaging time for each flux density sample is 12 hours.  The points marked by crosses denote 8\,GHz measurements reported by \citet{bach2007}. \label{fig:mainlc}}
\end{figure*}

Figure~\ref{fig:entirelc} demonstrates that the source exhibited variations on timescales of days and months to years during the period 2003 to 2005. During the
five day interval beginning 2005 July 20 the source exhibited a rapid, short timescale, 
increase in flux density from $1$ to $2.3\,$Jy, followed by a fall to
$1\,$Jy over the next five days. Over the following four days the
flux density recovered to a value of $1.6\,$Jy, close to the long term
mean flux density of the source (Figure~\ref{fig:mainlc}).

This event is atypical of the variability commonly observed in this
source and, indeed, all seven of the variable flat-spectrum quasars
regularly observed in the COSMIC program. The peak amplitude of the
flux density excursion exceeds the underlying mean by 44\% and occurs
on the exceedingly short timescale of only $\sim 4$\,days. This
behaviour contrasts markedly with the long-term intrinsic variability
evident in this source, as shown in Figure~\ref{fig:entirelc}. This
event is a one-off in the COSMIC dataset; no similar events have been
recorded in any of the other sources monitored in this program.

\section{Analysis}\label{sec:anyl}
\subsection{Origin of the Event}
The main clue as to the physical origin of this event lies in the fact
that it contains an excursion well {\it below} the average source flux
density.  Of the possible causes of this event -- scintillation
of a compact component, an intrinsic flare, microlensing by stars in an 
intervening galaxy, or an extreme scattering
event (ESE) in our Galaxy -- only an ESE is capable of affecting the entire source
brightness in such a way as to cause both the large positive and
negative fluctuations observed.

In the month prior to the event, the lightcurve is flat, indicating that
the flux density was dominated by quiescent emission originating from a
region of the source which, we argue here, is too large to exhibit
intrinsic variations on day timescales.  If variations on a timescale of
$\sim 40\,$days are to be attributed to intrinsic variability, causality
requires that the emission region encompass a size no larger than $\sim
0.03\,{\cal D}\,(1+z)^{-1}\,$pc, where ${\cal D}$ is the Doppler factor
associated with the motion of the emission region towards the observer.
However, the absence of interstellar scintillation in this source places a
lower limit on the source size which is incompatible with this limit.
Specifically, the absence of ISS implies an apparent angular size of $\ga
20\,\mu$as. Using cosmological parameters of $H_0$ = 71
kms$^{-1}Mpc^{-1}$, $\Omega_M$ = 0.27, $\Omega_{vac}$ = 0.73, the linear
size of the emitting region must therefore exceed $0.16\,{\cal D}$\,pc.
This size is not sufficiently small to allow causally-connected intrinsic
variations in the source on the timescale observed.

An interstellar scintillation-based explanation is unsatisfactory for the same
reason.  The onset of scintillation would require the appearance of a
new source component compact enough to exhibit ISS.  However, since
ISS can only affect the flux density of this new compact component,
and not the extended quiescent emission, the lightcurve can never dip
below the level of the quiescent emission.

Scattering due to the stellar wind of a nearby star along the line of 
sight to the source is also not a plausible explanation. We searched several 
Wolf-Rayet, O and B star catalogues through Vizier and found no nearby 
stars capable of generating bubbles of dense ionized matter with the 
properties required for this event.

Microlensing by stars in the intervening galaxies at $z=0.524$ and $z=0.851$ is similarly unable to explain the magnitude of the flux density excursion.  The $50\,\mu$as source size makes it impossible for microlensing to appreciably alter the entire source flux density because lensing only affects a region of size comparable to the Einstein radius.  Even for the closer of the two systems, the Einstein radius of a star of mass $m$ is only $\approx 1.56 \,(m/M_\sun)^{1/2}\,\mu$as (assuming $H_0 = 70\,$km\,s$^{-1}$Mpc$^{-1}$).  Futhermore, we note that the observed dip in the source flux density during the event is at variance with the magnification profile expected from a microlensing event.  A more detailed argument against the likelihood of microlensing in AO 0235$+$164 is presented in \citet{kraus1999}.

We are thus forced to conclude that only an event completely extrinsic
to the source is capable of affecting the source's emission, which is
too large to exhibit ISS and too large to be causally connected to an
intrinsic flare.  The most plausible explanation, therefore, is that
the flux density excursion is caused by an ESE.  We also note
that the lightcurve exhibits the large increase followed by a
subsequent sharp decrease which is characteristic of ESE events.  This
is consistent with the passing of the cloud across the line of sight
in which the radiation is first focused on to the observer and then,
as the observer passes the ESE boundary, the radiation is steered out
of the line of sight into the neighbouring focusing region.

\subsection{ESE Cloud Properties}
The duration of the ESE places constraints on the transverse length
scale of the ionising material.  The absence of data prior to and
following the high-amplitude variations places uncertainty on the
total length of the ESE event. We argue that the event had ceased by
day 947. This is supported by the 8 GHz data from Medicina \citep{bach2007}
that agree well with the Ceduna data where they overlap and, following
the period where rapid changes were observed, are consistent with the
flux density having returned to the underlying slow intrinsic
change. The earliest possible date for the commencement of the event
is day 916. Thus the maximum likely event duration is
$\Delta T=[947-916]$ days, while the minimum possible duration is 16
days assuming that the event is temporally symmetric about its peak, 
as is observed in other ESEs (e.g. \citet{fiedleretal87}). 

The transverse extent of the cloud is 
\begin{eqnarray}
r = 0.18 \left( \frac{v}{20\,{\rm km\,s}^{-1}} \right) 
\left( \frac{\Delta T}{16\,{\rm days}} \right)\, {\rm AU},
\end{eqnarray}
where $v$ is the speed of the cloud transverse to the line of sight,
plausibly in the range $10-50\,$km\,s$^{-1}$. Taking the uncertainties
in both $\Delta T$ and $v$ into account, the estimated linear
scale of the cloud is in the range 0.09-0.90\,AU.

Since the event significantly alters the entire flux density of the
source, the angular extent of the ionized cloud must be comparable to
or greater than the angle subtended by the core of the source. Space
VLBI observations of AO~0235+156 place an upper limit on the core size
at 4.8~GHz of $50\,\mu$as \citep{frey2000}. This, combined with the
favoured cloud size of 0.18\,AU requires a cloud distance of less than
3.6~kpc.

The electron density of an ESE cloud must be much higher than the
ambient medium. A simple estimate of the electron density is obtained by considering refraction by a cloud with a gaussian density profile: 
\begin{eqnarray}
n_e=n_0\exp \left[ \frac{-(r^2+z^2)}{2R_0^2} \right],
\end{eqnarray}
where $n_0$ is the maximum electron density, $R_0$ is the cloud
radius, $z$ is the distance along the line of sight, and $r$ is the distance from the cloud center transverse to the line of sight. The bending
angle, $\alpha$, of the cloud must  be comparable to the
angular size of the source so that,
\begin{eqnarray}
\Delta \theta \approx \alpha = \frac{1}{k}\frac{\partial \phi}{\partial r},
\end{eqnarray}
where $k$ is the wavenumber and $\phi$ is the phase delay imparted by the cloud. These
assumptions lead to a relationship between $n_0$ and $\alpha$ as a function of distance, $r$, from the cloud center: 
\begin{eqnarray}
\alpha= (2 \pi)^{1/2}\, n_0 \,r_e \lambda \, \frac{r}{R_0} \exp \left(- \frac{r^2}{2 R_0^2} \right),
\end{eqnarray}
where $r_e$ is the classical electron radius.  The maximum bending angle occurs at $r=R_0$; this requires a minimum cloud electron density of $4.1 \times 10^3\,$cm$^{-3}$, comparable to or higher than density estimates of previous ESE clouds \citep{fiedleretal87}.

The COSMIC program, to date, observed seven sources over a period
of four years, during which time one ESE was detected. Our observation
of a single event in 35 source years of monitoring is not inconsistent with
the ESE rate inferred by \citet{fiedleretal87}. This supports our
interpretation of this phenomenon as an ESE.

There have been a number of recent reports of short-duration transients at
high Galactic latitude \citep{bower2007,niinuma2007,matsumura2007}.  It is interesting to 
speculate whether short duration ESEs, similar to, or perhaps even more extreme than the event we have observed in
AO 0235+164 could be responsible for these events.  \citet{bower2007} used
archival VLA calibration observations to investigate the presence of radio
transients in a high-galactic latitude field for a period of about 20
minutes on average once every 7 days for more than 20 years.  They
identified 10 radio transient events where there is no persistent emission
stronger than a few $\mu$Jy.  However, in addition to these events their
data shows four cases where a source detected in the deep radio images was
detected in only a single epoch 20 minute observation.  Comparing the peak
and mean flux densities, the amplification observed in these events is of
the order of 2-5, larger than typical ESEs, but not implausibly so.  These
events may then be due to ESEs with similar timescales to the AO 0235$+$168
event ($<$ 20 days).  However, for the transients of \citet{bower2007}, and those
detected in the Nasu 1.4 GHz observations \citep{niinuma2007,matsumura2007} which have no detected persistent
radio emission the implied amplifications are of the order of 10-100 or more
and cannot plausibly explained by ESEs.

\section{Conclusion}

We observed a rapid 16 day timescale flux density variation in AO~0235+164 as part of the 6.7\,GHz COSMIC program. The presence of a flux density excursion below the underlying mean flux density of the source is consistent with an ESE cloud passing through the line of sight.

Assuming an ESE cloud velocity of 20\,km\,s$^{-1}$, we estimate the cloud to have a linear scale size ranging from 0.09 -- 0.90\,AU and distance less than 3.6\,kpc. We estimate the minimum electron density for the cloud to produce an ESE is $4 \times 10^3$\, cm$^{-3}$.

The detection and characterisation of this ESE required the dense flux density 
monitoring of the COSMIC program. The rapidity of this ESE  suggests that perhaps  other rapid ESEs may be overlooked in other monitoring programs due to undersampling of the events.  

\acknowledgements
This research has been supported by the ARC grant number DP0342500 at the University of Tasmania.
We thank Bev Bedson for her vital contribution to the operation of the Ceduna observatory and the 
COSMIC project. 
We thank Stefan Dieters for searching the archives of satellite based observatories for events coincident with the ESE detection. 
The National Radio Astronomy Observatory is a Facility of the National Science Foundation operated under co-operative agreement by Associated Universities, Inc.

\clearpage

\end{document}